# Empirical Evaluation of Integrated Trust Mechanism to Improve Trust in E-commerce Services

*Siddiqui Muhammad Yasir, Hyunsik Ahn

*Department of Robot System Engineering*

*Tongmyong University, Korea*

e-mail : *canuemail@gmail.com, hsahn@tu.ac.kr*

## Abstract

There are mostly two approaches to tackle trust management worldwide: *"Strong and crisp"* and *"Soft and Social"*. We analyze the impact of integrated trust mechanism in three different e-commerce services. The trust aspect is a dormant element between potential users and being developed expert or internet systems. We support our integration by preside over an experiment in controlled laboratory environment. The model selected for the experiment is a composite of policy and reputation based trust mechanisms and widely acknowledged in e-commerce industry. The integration between policy and trust mechanism was accomplished through mapping process, weakness of one brought to a close with the strength of other. Furthermore, experiment has been supervised to validate the effectiveness of implementation by segregating both integrated and traditional trust mechanisms in learning system.

## I. Introduction

The experiments are mostly executed as in laboratory controlled environment. We need to manipulate different variables in controlled environment. The purposed has been achieved within the laboratory environment. The objective of the empirical analysis is to determine the customer's satisfaction and trust on account holder in environment.

The experiment should be organized and planned in such a way and followed up, in order to control [1]. It is important to formulate the hypothesis before executing the experiment, under the laboratory environment. Our hypothesis may explained as, the trust level of new account holder is stayed on same level or its increased between proposed mechanism using our research questions.

The hypothesis statement is explained as: Null hypothesis, $H_0$ Support for new account holder the level of trust stayed same in comparison of proposed mechanism and traditional mechanisms [3].

Alternative hypothesis is explained as: $H_1$ Support the proposed mechanism increased the trust of new account holder in comparison of proposed and traditional mechanisms. There may be many selection variables can be involved within the organization or individual that entertained or measured within the organization [2].

## II. Background

There are two approaches widely popular to analyze the trust management in the industry. Personal and other entities experience in terms of feedback or rating is used to make a decision, called a reputation based trust mechanism [3]. The second trust mechanism is policy based trust mechanism where solid set of rules e.g. trusted certification authorities, signed certificates assigned by the system [1] [3].

There might be few cases where trust can't be fully achieved either by policy or reputation based trust mechanism. The purpose of this paper is to adopt one of the integrated approaches in learning system Using Indoor Object Recognition Based on Deep Learning and validate the results obtained by the experiment in controlled laboratory environment.

## III. Experiment Instruments

The design preparation is required before execution of experiment, conducting an experiment is not a simple task [2]. The design of experiment plays the role in success of execution and concludes meaningful outcomes of experiment. We learned with our experience the weak design can affect the execution and lead towards the unexpected results. The execution of experiment involved set of tasks, where each participant need to perform one by one after completing the tasks by the participant successfully a questionnaire asked to gather numerical data. The numerical data will be used to quantify proposed mechanism is beneficial and how much affected in the controlled environment.

### 3.1 Data Collection Procedure

The suggested design for experiment is *"One factor with more than two treatments"* and the Likert Scale is selected in order to collect the data from experiment [4]. Set of questions will be asked to the participants at the end of the experiment and answers will be scaled using Likert Scale. The Likert Scale is a psychometric scale commonly popular in research that employs questionnaires. The non-parametric test data using likert scale is suitable for hypothesis testing [5]. The non-parametric test Kruskal Wallis and Chi-2 is suggested in the book "Experimentation in Software Engineering". In order to perform the experiment for this particular research the data collection technique Likert Scale and Kruskal Wallis Test is selected for data analysis.

### 3.1.2 Instrumentation

The instrument of experiment is a planning phase of environment. A client machine is used to pretend real time environment as a server where http requests are being processed. A prototype of proposed integrated trust mechanism will be implemented on server machine. All the participants will engage with the client side machine and at the background our server will respond to the participants. To get the response of the experiment a questionnaire will be presented to the participants in order to collect the data and their experience about the proposed trust mechanism.

### 3.3 Experiment Variables

*"The variables are referred to as attributes/characteristics of an organization or individual that may be observed or measured within organization"* [2] [3]

### 3.3.1 Independent Variables

The proposed trust mechanism, Teradata and eBay were the treatments or independent variables to perform this experiment.

### 3.3.2 Dependent Variables

The user's decision can be a dependent variable just because of our experimental setup. The experiment was performed in controlled setup using our personal machines.

Students were selected and explained about the experiment and environment in order to achieve the same level of understating about the experiment.

The pilot client, server based application was developed to perform the experiment and generate the data for further analysis. Each participant performed the experiment individually, which was import to collect the original data for further analysis.

## IV. Experiment Execution

The objective of experiment execution is to evaluate proposed trust mechanism in controlled laboratory environment. The proposed trust mechanism can make right decisions and how they can be beneficial for the industry. Every participant have an introduction about the experiment just execution been started. The introduction was just for the understanding of participants about experiment. A combination of tasks performed by the participants during experiments execution as introduction was delivered to each participant just before the execution. Furthermore: a questionnaire was asked to the participants where every question needs to scale between 0 – 5 in numbers. The numerical data will be used with the Likert Scale and Kruskul Wallis test for data analysis, understanding the outcome of experiment.

### 4.1 Experiment Operation

The preparation is important to perform before executing the experiment along with participants. Set of tasks will be required to perform as per criteria for every participant, to what extent policy and trust based tasks effect on relation between participants as buyer or as seller. At the end each participant were asked to visit few products of interest after registration and analyze the trust opinions. Furthermore, trust values were evaluated in real time scenario using coding techniques in experiment setup [7]. At the end of the experiment, set of questions related to their tasks in experiment were asked to each participant for further calculation and analysis [8].

Questionnaires are related to the Enhance Security, Reputation Calculation, Cost, New Account Holder, Bad Image, and Customer Satisfaction. The analysis of collected data will be performed on sections mentioned above. The answers of questionnaires were divided on Likert Scale data analysis.

### 4.2 Statistical Analysis of Collected Data

The interpretation of collected data is carried out by given in numbers with hypothesis testing based on collected data from experiment. Kruskal Wallis test is a non-parametric test which is suitable for the situation where ANOVA normality assumptions cannot be applied [1]. As we are conducting experiment on the model design to measure the trust between two parties or more than two parties. The suitable test for comparing two or more than two treatments, involved one factor is ANOVA [1].

The mean calculation is explained below for further analysis of collected data.

**SUMMARY**

| Groups | Count | Min | Max | Sum | Mean | Median | Variance |
|---|---|---|---|---|---|---|---|
| Proposed Mechanism | 40 | 1 | 5 | 167 | 4.175 | 4 | 0.455769231 |
| Teradata | 40 | 1 | 4 | 89 | 2.225 | 2 | 0.845512821 |
| eBay | 40 | 1 | 3 | 81 | 2.025 | 2 | 0.486538462 |

Fig 1 Statistical summary of factors used in experiment

The statistical analysis is composed of equal size of sample data for each group. Greater no of samples or training data may produce different results. The calculated mean and average of proposed mechanism is clearly explaining that the trust of participants have increased significantly.

## V. Validation of Calculated Results

The hypothesis test will be performed in this section, on the measurements collection from the experiment. The kruskul wallis test is performed on each sample which is minimum five in numbers [6]. We need to calculate the means of three groups, a kruskul wallis test is handy in this situation. Now we can easily cast the interference among mean of three groups. The hypothesis testing is performed under one factor *"support for new account"*, where calculated degree of freedom *"K"* is representing the number of samples in experiment [6]. α typically set to 5% which is 0.05, and critical value is 5.99 used from chi square table using degree of freedom = 2 and α = 0.05 [6].

### 5.1 Calculate Mean of Ranks

The sum of ranks can be calculated by combining and arranging *"K"* samples and their assigned rank numbers. To avoid the *tie* in the case of data collection from Likert scale, mean value will be calculated of available rank numbers. The mean values calculated against ranks are explained below.

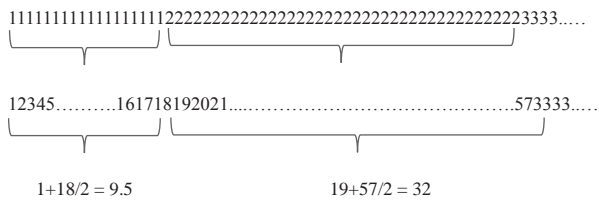

Fig 2 Mean values calculation example

The means values are calculated by organizing scale values and dividing them by first and last by number two. The calculated values are given below.

| Scale | Teradata | eBay | Proposed Mechanism | Mean Rank |
|---|---|---|---|---|
| No of 1s | 9 | 9 | 0 | 1+18/2 = 9.5 |
| No of 2s | 17 | 21 | 0 | 58+84/2 = 32 |
| No of 3s | 10 | 10 | 6 | 58+84/2 = 71 |
| No of 4s | 4 | 0 | 21 | 85+110/2 = 97.5 |
| No of 5s | 0 | 0 | 13 | 111+114/2 = 112.5 |
| n = 120 | n = 40 | n = 40 | n = 40 | |

Table 1: Mean values of Ranks

The procedure is already explained in [6], ties occurred in sample data to overcome this problem we need to calculate the mean values.

5.2 Hypothesis testing

The statistics test approximates chi-square distribution with the help of degree of freedom which is $K-1 = 2$ each $n_j = \{1,2,3\}$ in order to get valid approximation $n_j$ must be at least five. In our case $n=40$ greater than five so Kruskul Wallis test will be performed on collected data from experiment [9].

$$H = \frac{12}{N(N+1)} \sum_{i=1}^{g} n_i \bar{r}_{i.}^2 - 3(N+1)$$

The difference between all three treatments is considered significant. The value of $H$ is greater than 5.99 critical values [9].

The interpretation of results can be explained as the trust level of participant regarding support of new account holder's calculated mean is different between all three treatments.

The calculated $H$ value is greater than the critical value. Therefore: null hypothesis $H_0$ will be rejected in favor of alternative hypothesis. The hypothesis testing and descriptive analysis are explaining majority of participants are satisfied on the proposed trust mechanism.

## VI. Conclusion

In this research paper, we introduced and validated a technique to implement the trust mechanism between the relationship of new account holder and the system. The trust mechanism is validated by conducting the experiment in controlled environment to compare the identified factors between new account holder and the system. The paper we selected for implementation covers gap between policy and reputation based trust mechanisms. The proposed mechanism also facilitates few more factors e.g. enhanced mechanism for registration and reputation calculation.